\documentclass[12pt]{article}
\usepackage{subfigure}
\usepackage{amssymb,amsmath}
\usepackage{graphicx}
\usepackage{float}
\usepackage{color}
\usepackage{array,multirow}
\usepackage[font=footnotesize]{caption}
\usepackage[colorlinks=true
,urlcolor=blue
,citecolor=blue
,linkcolor=blue
,pagecolor=blue
,linktocpage=true
,pdfproducer=medialab
]{hyperref}
\usepackage[a4paper,width=16cm]{geometry}
\usepackage{cite}
\usepackage{placeins}
\usepackage[utf8]{inputenc}
\usepackage{cancel}
\usepackage{arydshln}
\makeatletter \renewcommand{\@dotsep}{10000} \makeatother
\usepackage{indentfirst}
\usepackage{rccol}
\usepackage{pifont}

\def\beq{\begin{equation}}
\def\eeq{\end{equation}}

\usepackage[nodisplayskipstretch]{setspace}

\begin{document}

\begin{titlepage}
\pagestyle{empty}

\vspace*{0.2in}
\begin{center}
{\Large \bf Higgs information in Split-SUSY at the LHC}\\[0.25cm]

\vspace{1cm}

{\bf {Surabhi Gupta\footnote{E-mail: sgupta2@myamu.ac.in} and Sudhir Kumar Gupta\footnote{E-mail: 
sudhir.ph@amu.ac.in} }}

\vspace{2pt}
	\begin{flushleft}
		{\em Department of Physics, Aligarh Muslim University, Aligarh, UP--202002, India} 
	\end{flushleft}
	
	\vspace{10pt}

\begin{abstract}
Information theory turns out to be an interesting tool for studying the consequences of Higgs observations to various new physics candidate theories by means of the information measure as the entropy of Higgs-Boson through its various detection modes at the Large Hadron Collider. The present article investigates the parameter space of a supersymmetric scenario where sfermions and one of the Higgs superfields are decoupled, while the gauginos, Higgsinos, and the remaining Higgs doublet are still allowed to be lighter. Our analysis reveals that this is quite a viable choice in the light of LHC discovery of a Higgs which resembles the SM Higgs-Boson and nothing else so far. While the supersymmetry breaking scale $M_S$ could be as high as $10^{11}$ GeV or so, the most preferred values of the $M_{S}$ and $\tan\beta$ are found to be around 3.6$\times10^7$ GeV and $41$ respectively,  
which is also consistent with the relic abundance of the neutralino dark matter. The corresponding value of neutralino ($ m_{\tilde\chi^{0}_{1}}$) LSP is estimated to be around 1.01 TeV. The preferred values of other parameters, namely, the Higgsino mass ($\mu$) and gaugino mass parameters ($M_1$ and $M_2$) are found to be about 1.05 TeV, 1.74 TeV, and 2.57 TeV, respectively.  
\end{abstract}
\end{center}
\end{titlepage}

\section{Introduction}
\label{sec:intro}

The observation of a Higgs-Boson at the LHC~\cite{Zyla:2020zbs} and the absence of any new resonances so far have posed a big challenge to isolate an appropriate candidate theory which could stabilise the Higgs mass against the radiative corrections. Supersymmetry (SUSY)~\cite{Martin:1997ns, Tata:1997uf, Drees:1996ca, Aitchison:2005cf, Fayet:2015sra, Djouadi:2005, Cane:2019ac, Allanchach:2019wrx} is one of the most prominent candidates that, in addition to provide a satisfactory mechanism for solving the issues related to Higgs mass, could also yield a dark matter candidate and explain the grand unification. However, the LHC observation of the Higgs-Boson with a mass beyond the expected value suggests much higher SUSY breaking scale $M_S$. Since the superpartners of fermions are not protected by any symmetry, their masses could grow as high as $M_S$, while the gauginos and Higgsinos could still be relatively much lighter as these are protected by the chiral symmetry. This splitting of masses of various sparticles, therefore, gives a split-supersymmetric (Split-SUSY) scenario~\cite{Wells:2004di,Arkani-Hamed:2004ymt,Giudice:2004tc,Arkani-Hamed:2004zhs}. Studies reveal that due to large values of $M_S$, Split-SUSY, in addition to explain the LHC observations, could also cure several problems of low-scale SUSY including the absence of FCNC contributions~\cite{Arkani-Hamed:2004ymt}, the grand unification~\cite{Arkani-Hamed:2004ymt,Giudice:2004tc}, the dark matter candidate as a neutralino~\cite{Giudice:2004tc,Wang:2013rba}, and proton decay in the case where R-parity is violated~\cite{Arkani-Hamed:2004ymt}.

The $M_S$ could be less than $10^{12}$ GeV by avoiding the dark matter candidature of gluinos. The mediation of decay of gluinos is carried out by the heavy squarks masses at the scale $M_S$, therefore gluinos can be long-lived or even their lifetime can be the age of the universe. Gluinos would be stable as their lifetime is the age of the universe when the $M_S$ is of the order of $10^{13}$ GeV~\cite{Arkani-Hamed:2004ymt,Arvanitaki:2005fa,Gambino:2005eh}. If gluino has a lifetime of more than a pico-second, the gluino would be hadronised to create a colour-singlet bound state comprising of gluino and quarks or gluons called “R-hadron” instead of decaying into a pair of quark-antiquark and the neutralino LSP~\cite{Fairbairn:2006gg,Arkani-Hamed:2004ymt}. Consequences of Split-SUSY have been widely studied in the context of colliders~\cite{Fairbairn:2006gg} as well as for cosmology~\cite{Arkani-Hamed:2004zhs,Demidov:2006zz,Demidov:2016wcv,Demidov:2017lzf}.

In the present article, we will explore the Split-SUSY scenario in greater detail using the information entropy of the observed Higgs-Boson assuming it to Standard Model (SM)~\cite{Djouadi:2005gi}. In Ref.~\cite{dEnterria:2012eip}, the study has been done by maximising the product of branching ratios of Higgs-Boson constructively used for measuring the preferable mass of SM Higgs-Boson that is well consistent with the LHC observed Higgs mass\cite{Zyla:2020zbs}. This approach links to the physical phenomena associated with the maximum possible decays. The branching ratios of Higgs-Boson constitute an information entropy, so using the process of the Maximum Entropy Principle (MEP) the precise mass of SM Higgs-Boson has been evaluated in Ref.~\cite{Alves:2014ksa}. Information entropy has been found to be quite successful in the investigation of new decay modes of the Higgs-Boson~\cite{Alves:2020cmr} and particles~\cite{Millan:2018fme,Llanes-Estrada:2017clj} at the LHC, axion mass evaluation~\cite{Alves:2017ljt}, and has also been recently applied in studying SUSY models~\cite{Gupta:2020whs,Gupta:2022psc}.

In our study, information entropy is used for investigating the Split-SUSY scenario. The information entropy of the Higgs-Boson~\cite{Alves:2014ksa} could be constructed by means of the branching ratios of the Higgs-Boson and it is maximised for a given Higgs mass. Further, the Higgs entropy could be used as a tool to predict the various sparticle masses at the LHC in light of experimental constraints from other experiments. In earlier studies, it has been shown that features of the Higgs-Boson in Split-SUSY resemble the SM Higgs-Boson, and therefore distinction between these may be quite challenging in experiments~\cite{Gupta:2005fq}.

The structure of the paper is organised as follows. In Section II, we discuss the Split-SUSY scenario and its existing Higgs-Boson. In Section III, we study the information entropy of the CP-even lightest Higgs-Boson. In Section IV, we analyse the Split-SUSY model in the context of information theory. In Section V, we summarise our findings.
 
\section{Split--SUSY and the Higgs--Boson}
In the Split-SUSY spectrum, masses of sfermions and one of the two Higgs doublets are present at a quite high scale $\simeq M_S$, in addition to SM contents, only gauginos, Higgsinos, and the lightest CP-even Higgs-Boson are found to be at the electroweak (EW) scale. 
Here, fermions are present at the EW scale that are protected against the radiative correction by the chiral symmetry, the unification of gauge couplings can be attained, and the evidence for the dark matter candidate as neutralino can also be obtained. These could be characterised by the most general renormalisable Lagrangian of the Split-SUSY, in terms of only the lightest CP-even Higgs-Boson (H), while heavy scalars confined at the $M_S$ are integrated out~\cite{Giudice:2004tc},
\beq \label{eq::1}
   \begin{split}
{\cal L}=m^2 H^\dagger H-\frac{\lambda}{2}\left( H^\dagger H\right)^2
-\left[h^u_{ij} {\bar q}_j u_i\epsilon H^* +h^d_{ij} {\bar q}_j d_iH
+h^e_{ij} {\bar \ell}_j e_iH \right. \\
+\frac{M_3}{2} {\tilde g}^A {\tilde g}^A +\frac{M_2}{2} 
{\tilde W}^a {\tilde W}^a +\frac{M_1}{2} {\tilde B} {\tilde B}
+\mu {\tilde H}_u^T\epsilon {\tilde H}_d \\
\left. +\frac{H^\dagger}{\sqrt{2}}\left({\tilde g}_u \sigma^a {\tilde W}^a 
+{\tilde g}_u^\prime {\tilde B} \right) {\tilde H}_u
+\frac{H^T\epsilon}{\sqrt{2}}\left(
-{\tilde g}_d \sigma^a {\tilde W}^a
+{\tilde g}_d^\prime {\tilde B} \right) {\tilde H}_d +{\rm h.c.}\right] ,
\end{split}
\eeq
where ${\tilde H}_{u,d}$, $\tilde
g$, $\tilde W$, and $\tilde B$ are Higgsinos, gluino, W-ino, and B-ino, respectively, while $\epsilon =i\sigma^2$, $\sigma^a$ denote the Pauli matrices, $\mu$ is the Higgsino mass parameter, ${\tilde g}_u$, ${\tilde g}_u^\prime$, ${\tilde g}_d$, and ${\tilde g}_d^\prime$ are gaugino couplings, and $M_1$, $M_2$, and $M_3$ are gaugino mass parameters corresponding to B-ino, W-ino, and gluino, respectively.

The strength of couplings at the $M_S$ associated with the effective theory in the Lagrangian of Eq.~\ref{eq::1}, can be evaluated by matching the interaction terms of Higgs doublets $H_u$ and $H_d$ of the SUSY Lagrangian, 

\beq
\begin{split}
{\cal L}_{\rm susy}=
-\frac{g^2}{8}\left( H_u^\dagger \sigma^a H_u + H_d^\dagger \sigma^a
H_d \right)^2
-\frac{g^{\prime 2}}{8}\left( H_u^\dagger H_u - H_d^\dagger  H_d
\right)^2 \\
+Y^u_{ij}H_u^T\epsilon {\bar u}_i q_j
-Y^d_{ij}H_d^T\epsilon {\bar d}_i q_j
-Y^e_{ij}H_e^T\epsilon {\bar e}_i \ell_j
 \\
-\frac{H_u^\dagger}{\sqrt{2}}\left( g \sigma^a {\tilde W}^a
+g^\prime {\tilde B} \right) {\tilde H}_u
-\frac{H_d^\dagger}{\sqrt{2}}\left(
g \sigma^a {\tilde W}^a
-g^\prime {\tilde B} \right) {\tilde H}_d +{\rm h.c.} 
\end{split}
\label{eq::2}
\eeq

The term $H=-\cos\beta \epsilon H_d^*+\sin\beta H_u$ in terms of two Higgs doublets acts as fine-tuning for acquiring small mass $m^2$. Now, the matching conditions about coupling constants, particularly at the $M_S$ in Eq.~\ref{eq::1} can be assessed by substituting $H_u\to \sin\beta H$ and $H_d\to \cos\beta \epsilon
H^*$ in Eq.~\ref{eq::2} as follows~\cite{Giudice:2004tc}
\beq
\lambda(M_S)= \frac{\left[ g^2(M_S)+g^{\prime 2}(M_S)        
\right]}{4} \cos^22\beta,
\label{eq::3}\\
\eeq
\beq
h^u_{ij}(M_S)= Y^{u*}_{ij}(M_S)\sin\beta , ~~~~~
h^{d,e}_{ij}(M_S)= Y^{d,e*}_{ij}(M_S)\cos\beta ,
\label{eq::4}
\eeq
\beq
{\tilde g}_u (M_S)= g (M_S)\sin\beta ,~~~~~
{\tilde g}_d (M_S)= g (M_S)\cos\beta ,
\label{eq::5}
\eeq
\beq
{\tilde g}_u^\prime (M_S)= g^\prime (M_S) \sin\beta ,~~~~~
{\tilde g}_d^\prime (M_S)= g^\prime (M_S)\cos\beta.
\label{eq::6}
\eeq
where $g$ and $g^\prime$ denote the gauge couplings, while $Y$'s represent the Yukawa couplings at the $M_S$ with two Higgs doublets and $\lambda$ considers the scalar self-coupling or the quartic coupling present in the theory having a single Higgs doublet.
$h^{(u,d,e)}$ are the Yukawa interactions of the existing Higgs doublet and can be evaluated from the matching conditions.

The only lightest CP-even Higgs termed as $h$ contributing to the low-energy effective Lagrangian of Eq.~\ref{eq::1} and its corresponding coupling can be achieved by putting $\beta-\alpha = \pi/2$ denoted as decoupling limit in the two-Higgs doublet Lagrangian of Eq.~\ref{eq::2}. The values of Yukawa and gauge couplings present at the low energy can be acquired via the evolution process from the scale $M_S$ using the matching conditions as given in Eqs~\ref{eq::3}--\ref{eq::6}. The mass of the lightest CP-even Higgs-Boson at the EW scale can be evaluated as
 \beq
m_h \sim \sqrt{\lambda} v, 
\label{eq::7}  
\eeq
where $v$ is the vacuum expectation value and $\lambda$ at the low energy is regulated by the logarithmically emphasised contribution received in the evolution process from the high scale $M_S$ using Eq~\ref{eq::3}.
 
\section{Information theory and Split--SUSY}
Shannon~\cite{shannon} describes entropy as an estimation tool of uncertainty associated with the information content. Information theory~\cite{jaynes:1957,thomas:2006} relies on probability theory in which each probability of an event of probability distribution inferred from Shannon's entropy (Eq. 2 of Ref.~\cite{Gupta:2020whs}) carries information, where information is considered to be the negative logarithm of the probability distribution. The maximum entropy of the system refers to the state of equilibrium with maximum uncertainty that provides maximum information, allowing MEP to predict the best value of the variable corresponding to the probability distribution. The information entropy (or Shannon’s entropy) with MEP is discussed in detail as well as related to the analysis of the Higgs-Boson under the CMSSM model~\cite{Gupta:2020whs} and also under the influence of NMFV over the CMSSM model~\cite{Gupta:2022psc}.

For including the MEP approach into our work, we need to calculate the Higgs entropy using Shannon's entropy concerning the branching fractions of the Higgs decay channels such as $h\rightarrow\gamma\gamma$, $h\rightarrow \gamma Z$, $h\rightarrow Z Z^*$, $h\rightarrow W W^*$, $h\rightarrow gg$, $h\rightarrow f\bar{f}$ with $f\in \{u, d, c, s, b, e, \mu, \tau \}$. Then by maximising the Higgs entropy, we observe that the Higgs mass is in good agreement with the measured Higgs mass at the LHC~\cite{Zyla:2020zbs}. 
For this, we assume an ensemble of $\cal N$-independent existing Higgs-Bosons at the EW scale in the light of information theory observed at the LHC could decay into above-mentioned allowed modes containing probabilities $p_{_j} (m_h)$ with its respective branching ratio $Br_j (m_h)$ as $p_{_j} (m_h) \equiv Br_j (m_h) =  \frac{\Gamma_j (m_h)}{\Gamma (m_h)}$, where $\Gamma_j (m_h)$ associates with the partial decay width of the $j^{th}$ decay mode of Higgs-Boson, while $\Gamma (m_h) = \sum_{j = 1}^{n_j}\Gamma_{j}(m_h)$ refers to the total decay width of all allowed decay modes of Higgs-Boson and $n_j$ is the total number of allowed Higgs-Boson decay modes.

The probability of the considered ensemble in the build of the multinomial distribution, emphasised in~\cite{Alves:2014ksa}, following each Higgs-Boson achieved its final state decaying to its possible decay modes can be specified by
\beq
\label{eq:8}
{\cal P}_{\{m{_j}\}}(m_h) = \frac{{\cal N}!}{m_1!...m_{n_j}!}\prod_{j = 1}^{n_j}{(p_{j} (m_h))}^{m_j}, 
\eeq
where $\sum^{n_j}_{j = 1} {Br}_{j} = 1 $, $ \sum_{j = 1}^{n_j}m_{j} = {\cal N}$, and the number of Higgs-Bosons decay particularly to $j^{th}$ mode is denoted as $m_{j}$.
Thus, the Shannon entropy of the ensemble under consideration is described as~\cite{Alves:2014ksa}   

\beq
\label{eq::9} 
S (m_h) = - \sum^{\cal N}_{\lbrace m_j \rbrace} {\cal P}_{\{m{_j}\}}(m_h) \ln {\cal P}_{\{m{_j}\}}(m_h).
\eeq
Further, an asymptotic expansion of the above-described information entropy would lead to a new form of entropy which is represented as~\cite{Alves:2014ksa}

\beq
\label{eq::10}
S (m_h) \simeq \frac{1}{2}\ln\left(\left(2\pi {\cal N} e\right)^{n_j -  1} \prod_{j = 1}^{n_j}{p_{j} (m_h)}\right) + \frac{1}{12 {\cal N}}\left( 3 n_j - 2 - \sum_{j = 1}^{n_j}{(p_{j} (m_h))}^{-1}\right)+ {\cal O}\left({\cal N}^{-2}\right).
\eeq

\section{Results and Discussions}
This section exposes the Split-SUSY analysis in the light of information theory. Information theory is a tool that requires only the branching ratios of the Higgs-Boson decays to determine the Higgs mass and can then be used to estimate the masses of sparticles effectively. For detailed analysis, we conduct a random scan concerning the free parameters described in the following range, 

\begin{itemize}
\item $\mu \in [0.1, 10] $ TeV,
\item $M_1 \in [0.1, 10] $ TeV,
\item $M_2 \in [0.1, 10]$ TeV,
\item $tan\beta \in [0.01, 70]$,
\item $M_S \in [10^4,10^8]$ GeV.
\end{itemize}
\begin{table}[h]
 \begin{centering}
    \tabcolsep 0.4pt
    \small
    \begin{tabular}{cccc}
    \hline
    \hline
    {Constraint}& {Observable}  & {Experimental~Value}& {Source}\\
       \hline 
LEP&$ m_{\tilde\chi^{0}_{1,2,3,4}}$ & $>$ 0.5 $m_Z$& \cite{Zyla:2020zbs}\\
&$ m_{\tilde\chi^{\pm}_{1,2}}$& $>$ 103.5 GeV& \cite{Zyla:2020zbs}\\
\hline
PO&$ BR(b \to s\gamma)$ & $(3.32\pm0.15)\times10^{-4}$&  \cite{Zyla:2020zbs,Amhis:2019ckw}\\
&$BR(B^0_s \to \mu^+\mu^-)$& $(3.0\pm0.4)\times10^{-9}$&  \cite{Zyla:2020zbs}\\
\hline
DM & $ \Omega_{\chi}h^{2}$ & $0.1197\pm0.0022$& \cite{Planck:2015fie}\\
   \hline
LHC-HIGGS &$ m_{h}$ & $125.1\pm 0.14$ GeV& \cite{Zyla:2020zbs}\\   
    \hline
    \end{tabular}
     \caption{\sf{Experimentally measured values of various observables used in our study.}}
      \label{tab:table1}
   \end{centering}
\end{table}
\begin{figure}[t]
\begin{centering}
\includegraphics[angle=0,width=1.0\linewidth,height=19em]{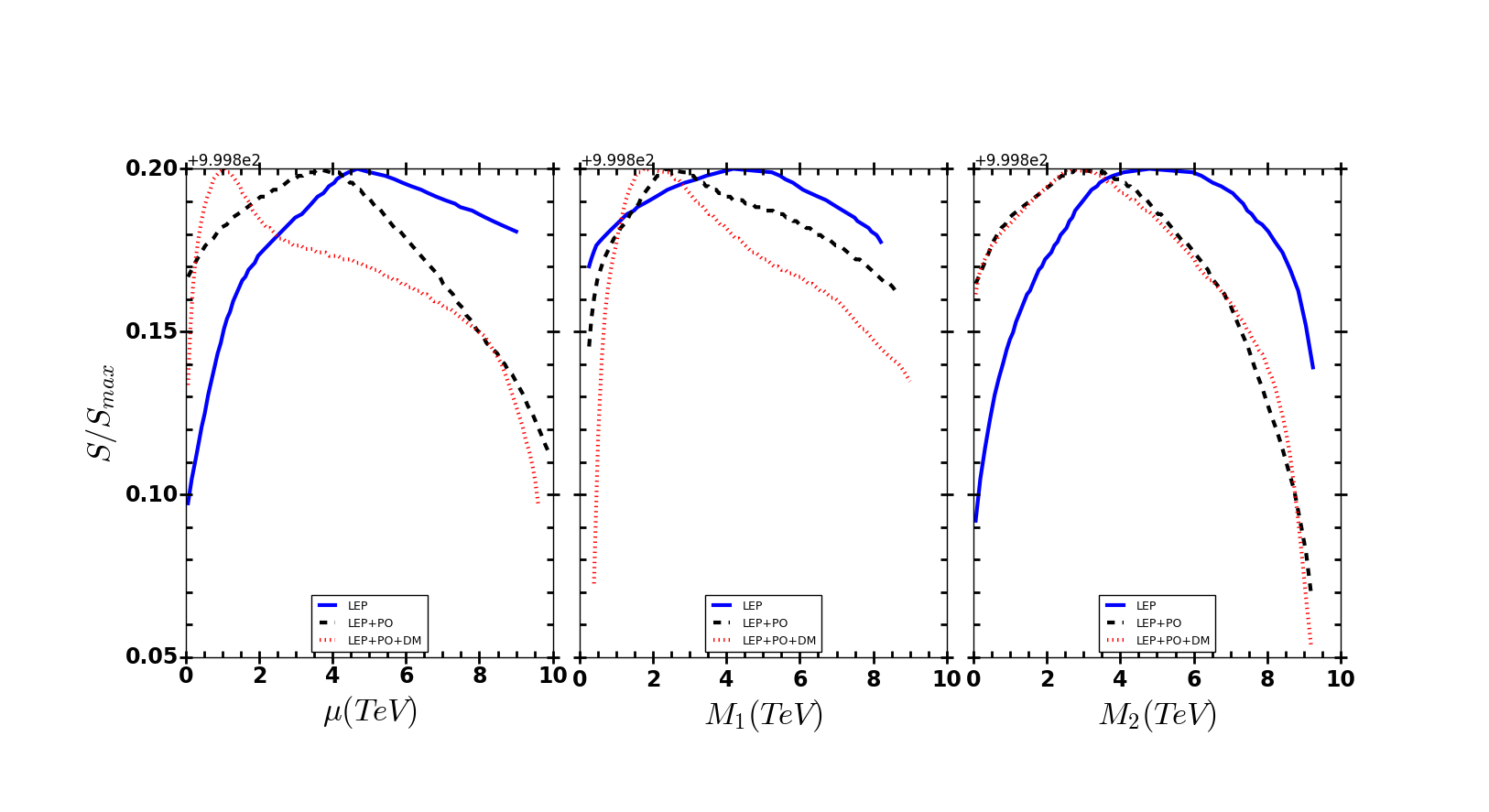}
\caption{\sf{Entropy vs Higgsino and gaugino mass parameters $\mu$ (left), $M_1$ (middle), and $M_2$ (right). The blue solid line represents the constraints from LEP data, the black dashed line contains the constraints from LEP data and B-Physics branching ratios, and the red dotted line indicates the constraints from LEP data, B-Physics branching ratios, and the relic abundance of the dark matter.}}
\label{fig:1}
\end{centering}
\end{figure}
\begin{figure}[t]
\begin{centering}
\includegraphics[angle=0,width=1.0\linewidth,height=19em]{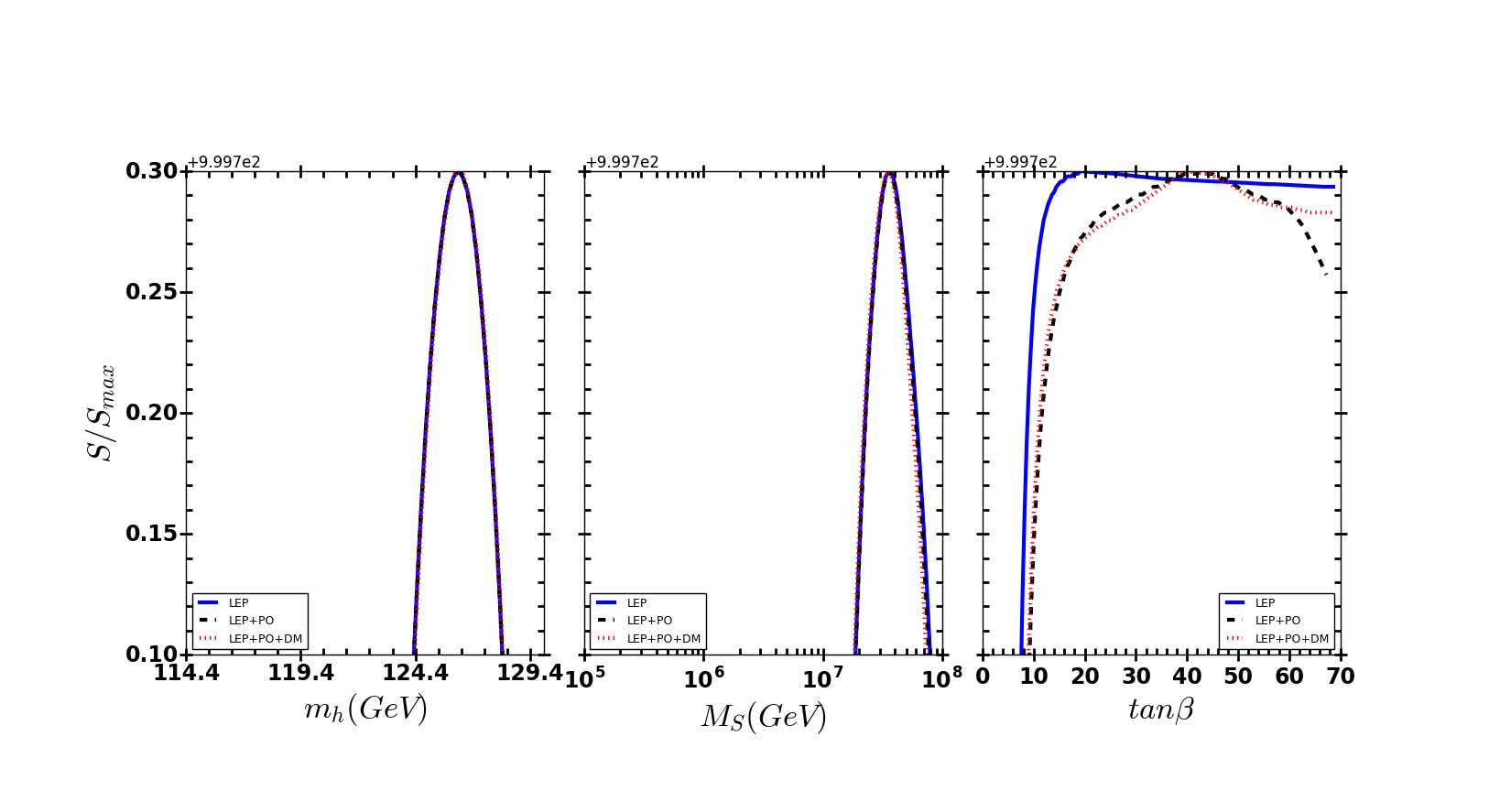}
\caption{\sf{Entropy vs (a) Higgs mass $m_h$ (left), (b) SUSY breaking scale $M_S$ (middle), and (c) $ tan\beta$ (right). The colour convention is similar to Figure~\ref{fig:1}.}}
\label{fig:2}
\end{centering}
\end{figure}
\noindent
where $\mu$ is the Higgsino mass parameter, $M_1$ and $M_2$ are gaugino mass parameters, $tan\beta$ ($\beta$ is the free parameter) constitutes a linear combination of two Higgs doublets where one Higgs doublet endure at the EW scale, while the other is at the $M_S$, acts as SM Higgs-doublet, instead of the ratio of vacuum expectation values of two Higgs doublets, and $M_S$ is the SUSY breaking scale which can be varied from $10^4$ GeV to $10^8$ GeV where the upper limit of the scale $M_S$ is restricted by the lifetime of the gluino as discussed in~\cite{Toharia:2005gm,Arvanitaki:2005fa,Gambino:2005eh}. Using the above-mentioned five free parameters, we use {\tt FlexibleSUSY}\cite{Athron:2014yba,Athron:2017fvs} to compute the masses of neutralinos, charginos, and Higgs-Boson, {\tt HDECAY}\cite{hdecay} to measure Higgs-Boson branching ratios, and {\tt SuperIso Relic}\cite{Arbey:2009gu,Arbey:2011,Arbey:2018msw} to assess observables such as $\triangle a_{\mu}$, $ BR(b \to s\gamma)$, $ BR(B^0_s \to \mu^+\mu^-)$, and $\Omega_{\chi}h^2$.

Information entropy is calculated using Eq.~\ref{eq::10} while considering an ensemble of $\cal N$-independent Higgs-Bosons at the EW scale which decay in its available detection modes such as $h\rightarrow\gamma\gamma$, $h\rightarrow \gamma Z$, $h\rightarrow Z Z^*$, $h\rightarrow W W^*$, $h\rightarrow gg$, $h\rightarrow f\bar{f}$ with $f\in \{u, d, c, s, b, e, \mu, \tau \}$. Then Higgs entropy employs for precise estimation of masses of the Higgsinos, gauginos, neutralinos, and charginos. Since the sfermion masses are ultra-heavy around the scale $M_S$, it has been realised that the additional contribution is provided by the charginos present in the loops for the Higgs decays, particularly in $h\rightarrow\gamma\gamma$ and $h\rightarrow \gamma Z$. Furthermore, the presence of the chargino loops in Higgs decays make Higgs existing at the EW scale distinguished from the SM Higgs~\cite{Gupta:2005fq}.

\begin{figure}[t]
\begin{centering}
\includegraphics[angle=0,width=1.0\linewidth,height=24em]{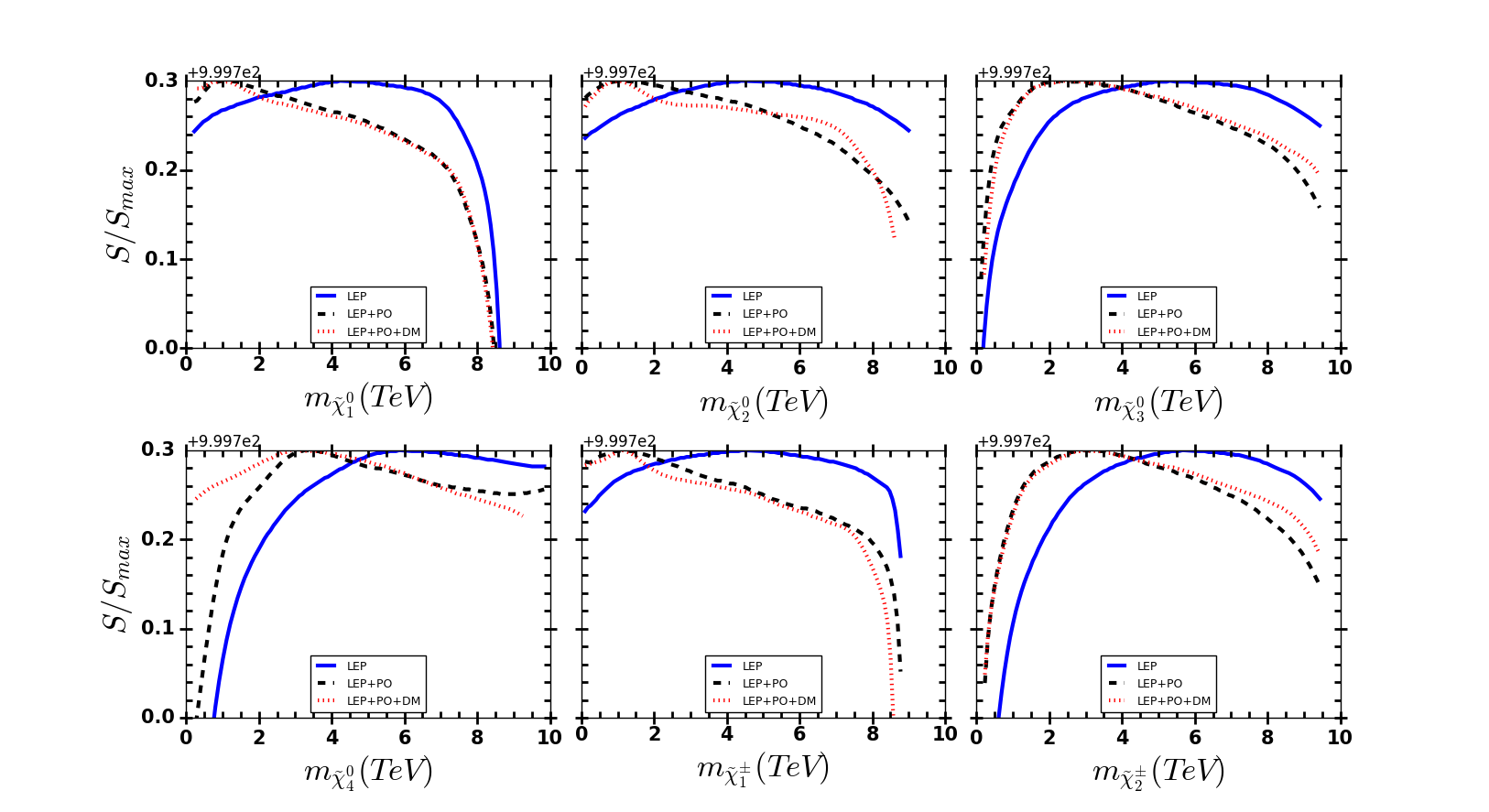}
\caption{\sf{Entropy vs neutralino and chargino masses for various constraints. The colour convention is similar to Figure~\ref{fig:1}.}}
\label{fig:3}
\end{centering}
\end{figure}

Information entropy depends only on Higgs mass $m_h$ following the marginalisation over all other parameters of our model and scaling it with a normalisation factor $1/S_{max}$. Thereafter, the parameter space has been imposed with constraints from LEP data on neutralino and chargino masses which are described as $ m_{\tilde\chi^{0}_{1,2,3,4}} >$ 0.5 $m_Z$ and $ m_{\tilde\chi^{\pm}_{1,2}} >$ 103.5 GeV, respectively, B-Physics branching ratios i.e. $ BR(b \to s\gamma)$, $ BR(B^0_s \to \mu^+\mu^-)$, dark matter relic density $\Omega_{\chi}h^2$, and the constraint on the Higgs mass from the LHC ($ m_h = $ 125.10 $\pm$ 0.14 GeV) at 2.5$\sigma$ confidence level as given in Table~\ref{tab:table1}. In our work, findings are exhibited in Figures~\ref{fig:1}--\ref{fig:3} in view of the following constraints (a) LEP, (b) LEP$+$PO, and (c) LEP$+$PO$+$DM. In Figures~\ref{fig:1}--\ref{fig:3}, the blue solid line includes the bounds on LEP data which provide the minimum mass limit on neutralinos and charginos, the black dashed line symbolises constraints from LEP data and B-Physics branching ratios, and the red dotted line illustrates constraints from LEP data, B-Physics branching ratios, and the relic abundance of the neutralino dark matter $\Omega_{\chi}h^2$.

We explore the spectrum of Split-SUSY in consideration of the information theory. We present the plots to show the variation of information entropy with $\mu$ (left), $M_1$ (middle), and $M_2$ (right) in Figure~\ref{fig:1}, with $m_h$ (left), $M_S$ (middle), and $\tan\beta$ (right) in Figure~\ref{fig:2}, and with neutralinos and charginos in Figure~\ref{fig:3}. Here, the Higgs mass is evaluated in the following way, considering particular values of $tan\beta$ and gauge couplings as boundary conditions at the $M_S$ in Eq.~\ref{eq::3} we can estimate $\lambda$ at the $M_S$. Then the RGEs can be solved to get $\lambda$ at the EW scale evolved down from the scale $M_S$, continuing the iterative process until the convergence has been reached. The Higgs mass at the EW scale can be obtained by putting $\lambda$ at the EW scale in Eq.~\ref{eq::7}. The combined ATLAS and CMS experimentally observed value of the Higgs mass at the LHC, $ m_h = $ 125.10 $\pm$ 0.14 GeV~\cite{Zyla:2020zbs}.

The preferable values of $\mu$, $M_1$, $M_2$, $M_S$, and $\tan\beta$ with respect to maximum entropy in constraints of LEP data are 4.67 TeV, 4.18 TeV, 4.79 TeV, 3.6$\times10^7$ GeV, and 19.1, respectively. The values of the above-mentioned parameters turn out to be 3.59 TeV, 2.3 TeV, 2.79 TeV, 3.6$\times10^7$ GeV, and 42.4, respectively, in constraints of LEP data and B-Physics branching ratios, while in concern of all applied constraints of LEP data, B-Physics branching ratios, and the relic abundance of the neutralino dark matter these values correspond to 1.05 TeV, 1.74 TeV, 2.57 TeV, 3.6$\times10^7$ GeV, and 41, respectively. In this scenario, the parameter space contains masses of the Higgsino and gaugino parameters to be of the TeV scale. The mass value of the lightest CP-even Higgs-Boson at the EW scale is predicted by our approach which is likely to be SM. The other Higgses and scalar sector are ultra-heavy found nearly at the scale $M_S$ and are decoupled from the EW scale. In addition, this scenario could also naively predict the dark matter candidate. The corresponding values for the aforementioned parameters are found to be 1.77 TeV, 7.27 TeV, 0.55 TeV, 5.9$\times10^6$ GeV, and 41.6, respectively, after including the LHC constraint on the mass of the Higgs-Boson as listed in Table~\ref{tab:table1}.

The preferable values associated with Higgs mass $m_h$, the lightest neutralino $ m_{\tilde\chi^{0}_{1}}$, and the lighter chargino $ m_{\tilde\chi^{\pm}_{1}}$ are 126.3 GeV, 4.23 TeV, and 4.41 TeV, respectively, including constraints from LEP data, whereas these correspond to 126.2 GeV, 1.04 TeV, and 1.09 TeV, respectively, in favour of constraints from LEP data and B-physics branching ratios. The values corresponding to the aforementioned parameters including constraints from LEP data, B-physics branching ratios, and the relic abundance of the dark matter are 126.3 GeV, 1.01 TeV, and 1.13 TeV, respectively. These after incorporating the LHC constraint on the mass of the Higgs-Boson turn out to be 125.45 GeV, 0.532 TeV, and 0.533 TeV, respectively. The masses of sparticles for the aforementioned parameters are listed in Table~\ref{tab:table3}. From Table~\ref{tab:table3}, it is to be noted that the neutralino LSP could have a mass of about 532 GeV while the mass of the gauginos could be upto about 7.3 TeV.
    
 \begin{table}[h!]
  \begin{center}
    \small
    \begin{tabular}{ccccc} 
      \hline
      \hline
        \multirow{2}{*}{Parameter}& \multicolumn{3}{c}{Constraints}& \\
      \cline{2-4} 
      & {LEP} & {LEP $+$ PO}& {LEP $+$ PO $+$ DM} \\
       \hline
       \hline
      $\mu$ &4.67&3.59&1.05&\\
      $ M_1$  &4.18&2.30&1.74&\\
      $ M_2$  &4.79&2.79&2.57&\\
      $ M_S$  &3.6$\times10^7$&3.6$\times10^7$&3.6$\times10^7$&\\
      $ tan\beta$  &19.1&42.4&41&\\
           \hline
      $m_h$&126.3&126.2&126.3&\\
      \hline       
      $ m_{\tilde\chi^{0}_{1}}$ &4.23&1.04&1.01&\\
      $ m_{\tilde\chi^{0}_{2}}$   &4.39&1.13&1.04&\\
       $ m_{\tilde\chi^{0}_{3}}$ &5.31&2.30&2.39&\\
      $ m_{\tilde\chi^{0}_{4}}$  &5.85&3.26&2.89&\\
      $ m_{\tilde\chi^{\pm}_{1}}$  &4.41&1.09&1.13&\\
      $ m_{\tilde\chi^{\pm}_{2}}$  &5.66&2.80&2.92&\\
                 \hline
      \hline
    \end{tabular}
    \caption{\sf{Split-SUSY spectrum with respect to maximum entropy. All masses are in TeV except $ M_S$ and $ m_h$ are in GeV.}}
    \label{tab:table2}
 \end{center}
\end{table}


 
 \begin{table}[h!]
  \begin{center}
    \small
    \begin{tabular}{lcccc} 
      \hline
      \hline
        {Parameter}& {Value}& \\
       \hline
       \hline
      $\mu$ &1.77&\\
      $ M_1$  &7.27 &\\
      $ M_2$  & 0.55&\\
      $ M_S$ (GeV) &5.9$\times10^6$&\\
      $ tan\beta$  &41.6&\\
      \hline
      $ m_{\tilde\chi^{0}_{1}}$ &0.532&\\
      $ m_{\tilde\chi^{0}_{2}}$  &1.73&\\
       $ m_{\tilde\chi^{0}_{3}}$ &1.73&\\
      $ m_{\tilde\chi^{0}_{4}}$  &7.29&\\
      $ m_{\tilde\chi^{\pm}_{1}}$ &0.533&\\
      $ m_{\tilde\chi^{\pm}_{2}}$  &1.73&\\
                 \hline
      \hline
    \end{tabular}
    \caption{\sf{Allowed Split-SUSY parameters and the corresponding spectrum with respect to maximum entropy after taking into account the LHC constraint on the mass of the Higgs-Boson. All masses are in TeV except stated otherwise.}}
    \label{tab:table3}
 \end{center}
\end{table}


\section{Summary}
In this study, we have investigated the Split-SUSY spectrum in the light of Higgs-Boson mass at the LHC. We have evaluated the entropy of Higgs-Boson concerning branching ratios of its allowed decay channels over the defined range of Split-SUSY parameter space against the experimental constraints from LEP data, B-physics branching ratios, the relic abundance of the dark matter, and the LHC constraint on the mass of the Higgs-Boson. Further, our study reveals that the Higgs entropy is capable of estimating the values of the rest of the parameters. We have presented our results in Figures~\ref{fig:1}--\ref{fig:3}. The variation of information entropy with respect to $\mu$ (left), $M_1$ (middle), and $M_2$ (right) are exhibited in Figure~\ref{fig:1}. Considering $\beta$ as a free parameter in our study, the information entropy is plotted with $m_h$ (left), $M_S$ (middle), and $\tan\beta$ (right) as shown in Figure~\ref{fig:2}. In Figure~\ref{fig:3}, we have presented the plots of information entropy with neutralinos and charginos.

Using our approach, the associated values for $\mu$, $M_1$, $M_2$, $M_S$, $\tan\beta$, $m_h$, the lightest neutralino $ m_{\tilde\chi^{0}_{1}}$, and the lighter chargino $ m_{\tilde\chi^{\pm}_{1}}$ are expected to be 4.67 TeV, 4.18 TeV, 4.79 TeV, 3.6$\times10^7$ GeV, 19.1, 126.3 GeV, 4.23 TeV, and 4.41 TeV, respectively, for LEP data constraints. After taking into account the constraints on LEP data and B-physics branching ratios, the values of aforementioned parameters correspond to 3.59 TeV, 2.3 TeV, 2.79 TeV, 3.6$\times10^7$ GeV, 42.4, 126.2 GeV, 1.04 TeV, and 1.09 TeV, respectively, whereas these values change to 1.05 TeV, 1.74 TeV, 2.57 TeV, 3.6$\times10^7$ GeV, 41, 126.3 GeV, 1.01 TeV, and 1.13 TeV, respectively, when the constraint of the relic abundance of the dark matter is also included. It is to be noted that the observed Higgs mass is in good agreement with the measured Higgs mass at the LHC~\cite{Zyla:2020zbs}. Our study also reveals that the lightest neutralino LSP should have a mass around 1.01 TeV including the constraints from LEP data, B-Physics branching ratios, and the relic abundance of the neutralino dark matter. After taking into account the constraint of experimental Higgs mass at 2.5$\sigma$ confidence level along with the above-mentioned constraints. The corresponding values for ($\mu$, $M_1$, $M_2$, $M_S$, $\tan\beta$, $ m_{\tilde\chi^{0}_{1}}$, $ m_{\tilde\chi^{0}_{2}}$, $ m_{\tilde\chi^{0}_{3}}$, $ m_{\tilde\chi^{0}_{4}}$, $ m_{\tilde\chi^{\pm}_{1}}$ and $ m_{\tilde\chi^{\pm}_{2}}$) are observed to be about (1.77 TeV, 7.27 TeV, 0.55 TeV, 5.9$\times10^6$ GeV, 41.6, 0.532 TeV, 1.73 TeV, 1.73 TeV, 7.29 TeV, 0.533 TeV, and 1.73 TeV), respectively as listed in Table~\ref{tab:table3}. 

\section*{Acknowledgement}
This work was supported in part by University Grant Commission under a Start-Up Grant no. F30-377/2017 (BSR). We thank Apurba Tiwari for useful discussions.
We acknowledge the avail of computing facility at the DST Computational lab, AMU, Aligarh, India during the initial work.



\begin{thebibliography}{44}


\bibitem{Zyla:2020zbs}
P.~A.~Zyla {\it et al.} [Particle Data Group], 
Prog.\ Theor.\ Exp.\ Phys.\ {\bf 2020}, no.8, 083C01 (2020).


\bibitem{Tata:1997uf}
X.~Tata,
[arXiv:hep-ph/9706307 [hep-ph]].


\bibitem{Martin:1997ns}
S.~P.~Martin,
Adv.\ Ser.\ Direct.\ High Energy Phys.\ {\bf 21}, 1-153 (2010),
[arXiv:hep-ph/9709356 [hep-ph]].


\bibitem{Drees:1996ca}
M.~Drees,
[arXiv:hep-ph/9611409 [hep-ph]].


\bibitem{Fayet:2015sra}
P.~Fayet,
Adv.\ Ser.\ Direct.\ High Energy Phys.\ {\bf 26}, 397-454 (2016),
[arXiv:1506.08277 [hep-ph]].


\bibitem{Cane:2019ac}
A.~Canepa,
Rev.\ Phys.\ {\bf 4}, 100033 (2019).


\bibitem{Allanchach:2019wrx}
B.~C.~Allanchach,
CERN Yellow Rep.\ School Proc.\ {\bf 6}, 113-144 (2019).


\bibitem{Aitchison:2005cf}
I.~J.~R.~Aitchison,
[arXiv:hep-ph/0505105 [hep-ph]].


\bibitem{Djouadi:2005} 
A.~Djouadi,
Phys.\ Rept.\ {\bf 459}, 1-241 (2008),
[arXiv:hep-ph/0503173 [hep-ph]].


\bibitem{Wells:2004di}
J.~D.~Wells,
Phys.\ Rev.\ D {\bf 71}, 015013 (2005),
[arXiv:hep-ph/0411041 [hep-ph]].


\bibitem{Arkani-Hamed:2004ymt}
N.~Arkani-Hamed and S.~Dimopoulos,
JHEP {\bf 06}, 073 (2005),
[arXiv:hep-th/0405159 [hep-th]].


\bibitem{Giudice:2004tc}
G.~F.~Giudice and A.~Romanino,
Nucl.\ Phys.\ B {\bf 699}, 65-89 (2004).
[erratum: Nucl.\ Phys.\ B {\bf 706}, 487-487 (2005)]
[arXiv:hep-ph/0406088 [hep-ph]].



\bibitem{Arkani-Hamed:2004zhs}
N.~Arkani-Hamed, S.~Dimopoulos, G.~F.~Giudice and A.~Romanino,
Nucl.\ Phys.\ B {\bf 709}, 3-46 (2005),
[arXiv:hep-ph/0409232 [hep-ph]].


\bibitem{Wang:2013rba}
F.~Wang, W.~Wang and J.~M.~Yang,
Eur.\ Phys.\ J.\ C {\bf 74}, no.10, 3121 (2014),
[arXiv:1310.1750 [hep-ph]].


\bibitem{Arvanitaki:2005fa}
A.~Arvanitaki, C.~Davis, P.~W.~Graham, A.~Pierce and J.~G.~Wacker,
Phys.\ Rev.\ D {\bf 72}, 075011 (2005),
[arXiv:hep-ph/0504210 [hep-ph]].


\bibitem{Gambino:2005eh}
P.~Gambino, G.~F.~Giudice and P.~Slavich,
Nucl.\ Phys.\ B {\bf 726}, 35-52 (2005),
[arXiv:hep-ph/0506214 [hep-ph]].


\bibitem{Fairbairn:2006gg}
M.~Fairbairn, A.~C.~Kraan, D.~A.~Milstead, T.~Sjostrand, P.~Z.~Skands and T.~Sloan,
Phys.\ Rept.\ {\bf 438}, 1-63 (2007),
[arXiv:hep-ph/0611040 [hep-ph]].



\bibitem{Demidov:2006zz}
S.~V.~Demidov and D.~S.~Gorbunov,
JHEP {\bf 02}, 055 (2007),
[arXiv:hep-ph/0612368 [hep-ph]].


\bibitem{Demidov:2016wcv}
S.~V.~Demidov, D.~S.~Gorbunov and D.~V.~Kirpichnikov,
JHEP {\bf 11}, 148 (2016),
[erratum: JHEP \textbf{08}, 080 (2017)]
[arXiv:1608.01985 [hep-ph]].


\bibitem{Demidov:2017lzf}
S.~V.~Demidov, D.~S.~Gorbunov and D.~V.~Kirpichnikov,
Phys.\ Lett.\ B {\bf 779}, 191-194 (2018),
[arXiv:1712.00087 [hep-ph]].


\bibitem{Djouadi:2005gi}
A.~Djouadi,
Phys.\ Rept.\ {\bf 457}, 1-216 (2008), 
[arXiv:hep-ph/0503172 [hep-ph]].



\bibitem{dEnterria:2012eip}
D.~d'Enterria,
``On the Gaussian peak of the product of decay probabilities of the standard model Higgs boson at a mass $m_H\sim$125 GeV,''
[arXiv:1208.1993 [hep-ph]].


\bibitem{Alves:2014ksa}
A.~Alves, A.~G.~Dias and R.~da Silva,
Physica A {\bf 420}, 1-7 (2015),
[arXiv:1408.0827 [hep-ph]].


\bibitem{Alves:2020cmr}
A.~Alves, A.~G.~Dias and R.~da Silva,
Nucl.\ Phys.\ B {\bf 959}, 115137 (2020),
[arXiv:2004.08407 [hep-ph]].


\bibitem{Llanes-Estrada:2017clj}
F.~J.~Llanes-Estrada, P.~C.~Millan, A.~Porras Riojano, E.~M.~Sánchez García and M.~Á.~García Ferrero,
PoS {\bf EPS-HEP2017}, 740 (2017),
[arXiv:1710.01286 [hep-ph]].


\bibitem{Millan:2018fme}
P.~Carrasco Millán, M.~Á.~García-Ferrero, F.~J.~Llanes-Estrada, A.~Porras Riojano and E.~M.~Sánchez García,
Nucl.\ Phys.\ B {\bf 930}, 583-596 (2018),
[arXiv:1802.05487 [hep-ph]].


\bibitem{Alves:2017ljt}
A.~Alves, A.~G.~Dias and R.~Silva,
Braz.\ J.\ Phys.\ {\bf 47}, no.4, 426-435 (2017),
[arXiv:1703.02061 [hep-ph]].


\bibitem{Gupta:2020whs}
S.~Gupta and S.~Kumar Gupta,
Nucl.\ Phys.\ B {\bf 965}, 115336 (2021),
[arXiv:2008.00415 [hep-ph]].


\bibitem{Gupta:2022psc}
S.~Gupta and S.~K.~Gupta,
Nucl.\ Phys.\ B {\bf 984}, 115942 (2022),
[arXiv:2205.00173 [hep-ph]].


\bibitem{Gupta:2005fq}
S.~K.~Gupta, B.~Mukhopadhyaya and S.~K.~Rai,
Phys.\ Rev.\ D {\bf 73}, 075006 (2006),
[arXiv:hep-ph/0510306 [hep-ph]].



\bibitem{shannon}
C.~E.~Shannon, 
The Bell Syst.\ Tech.\ J.\ {\bf 27}, 379-423 (1948).


\bibitem{jaynes:1957}
E.~T.~Jaynes, 
Phys.\ Rev.\ {\bf 106}, 620-630 (1957).

 
\bibitem{thomas:2006}
T.~M.~Cover and J.~A.~Thomas, 
Second Edition, Wiley-Interscience (2006). 


\bibitem{Toharia:2005gm}
M.~Toharia and J.~D.~Wells,
JHEP {\bf 02}, 015 (2006),
[arXiv:hep-ph/0503175 [hep-ph]].



\bibitem{Athron:2014yba}
P.~Athron, J.~h.~Park, D.~St\"ockinger and A.~Voigt,
Comput.\ Phys.\ Commun.\ {\bf 190}, 139-172 (2015),
[arXiv:1406.2319 [hep-ph]].


\bibitem{Athron:2017fvs}
P.~Athron, M.~Bach, D.~Harries, T.~Kwasnitza, J.~h.~Park, D.~St\"ockinger, A.~Voigt and J.~Ziebell,
Comput.\ Phys.\ Commun.\ {\bf 230}, 145-217 (2018),
[arXiv:1710.03760 [hep-ph]].


\bibitem{hdecay}
Abdelhak Djouadi, Jan Kalinowski, Margarete Mühlleitner and Michael Spira,
Comput.\ Phys.\ Commun. {\bf 238}, 214-231 (2019),

\bibitem{Arbey:2009gu}
A.~Arbey and F.~Mahmoudi,
Comput.\ Phys.\ Commun. {\bf 181}, 1277-1292 (2010),
[arXiv:0906.0369 [hep-ph]].


\bibitem{Arbey:2011}
Comput.\ Phys.\ Commun. {\bf 182}, 1582-1583 (2011),


\bibitem{Arbey:2018msw}
A.~Arbey, F.~Mahmoudi and G.~Robbins,
Comput.\ Phys.\ Commun.\ {\bf 239}, 238-264 (2019),
[arXiv:1806.11489 [hep-ph]].


\bibitem{Amhis:2019ckw}
Y.~S.~Amhis {\it et al.} [HFLAV],
Eur.\ Phys.\ J.\ C {\bf 81}, no.3, 226 (2021),
[arXiv:1909.12524 [hep-ex]].


\bibitem{Planck:2015fie}
P.~A.~R.~Ade \textit{et al.} [Planck],
Astron.\ Astrophys.\ {\bf 594}, A13 (2016),
[arXiv:1502.01589 [astro-ph.CO]].


\end{thebibliography}
\end{document}